\documentclass[prb,10pt,twocolumn,superscriptaddress,footnoteinbib]{revtex4}
\usepackage{amsmath}
\usepackage{latexsym}
\usepackage{amssymb}
\usepackage{graphics,epstopdf}
\usepackage[colorlinks=true, citecolor=blue, urlcolor=blue ]{hyperref}
\usepackage{epsf,graphics,graphicx}

\date{\today}

\begin{document}

\title{Spin filtration in a single-stranded antiferromagnetic helix with slowly varying disorder: Higher order electron hopping}

\author{Suparna Sarkar}

\email{suparna@jncasr.ac.in}

\affiliation{Theoretical Sciences Unit, Jawaharlal Nehru Centre for Advanced Scientific Research, Bangalore-560 064, India}

\author{Santanu K. Maiti}

\email{santanu.maiti@isical.ac.in}

\affiliation{Physics and Applied Mathematics Unit, Indian Statistical Institute, 203 Barrackpore Trunk Road, Kolkata-700 108, India}

\author{David Laroze}

\email{dlarozen@uta.cl}

\affiliation{Instituto de Alta Investigaci\'{o}n, Universidad de Tarapac\'{a}, Casilla 7D, Arica, Chile}

\begin{abstract}

This work explores spin filtration in a helical magnetic system within a tight-binding framework, where neighboring magnetic moments are 
aligned antiparallel. The helix experiences a slowly-varying diagonal disorder, following a cosine form, which creates a finite energy 
mismatch between up and down spin channels. Unlike earlier studies that relied on external electric fields, this investigation demonstrates 
that disorder alone can achieve high spin filtration, even at low bias and high temperatures. Higher-order electron hopping in the helix 
leads to a non-uniform energy level distribution, facilitating favorable spin filtration, sometimes reaching $100\%$. The interplay between
higher-order hopping and atypical disorder may enable selective spin transmission through various antiferromagnetic helices, potentially 
opening new avenues for functional elements.

\end{abstract}

\maketitle

\section{Introduction}

The role of disorder in physical systems has always been a fascinating subject, leading to the exploration of various peculiar phenomena.
It started in the 1950s with the pioneering work of P. W. Anderson~\cite{anderson} which showed that for a one-dimensional system with
uncorrelated (random) disorder and nearest-neighbor electron hopping all states become exponentially localized, irrespective of disorder
strength~\cite{random1,random2,random3,random4,random5,random6}. The existence of zero critical disorder strength sometimes makes random
disordered systems quite trivial, and here no one encountered any mobility edge (ME) that separates localized states from the extended
ones~\cite{mp1,mp2}. The situation becomes interesting once we impose a specific correlation~\cite{cor1,cor2,cor3,cor4,cor5} in disorder
distribution. Among many models, the Aubry-Andr\'{e}-Harper (AAH)~\cite{aah1,aah2} model is a well-established correlated 
disordered system that exhibits several atypical behavior and has been widely studied in the past. Most commonly diagonal AAH model is
taken into account where site energies are described in the form: $W \cos(2 \pi b n)$, where $W$ measures the disorder strength, $n$
represents the site index and $b$ is an incommensuration parameter~\cite{zil}. Imposing this specific correlation among site energies, it is 
possible to set a finite critical disorder strength beyond which the system goes to the localized regime, though in this case, one still
cannot find any ME. The existence of MEs is quite significant when we think about selective particle transfer through a material. This
issue has been solved by including higher order electron hopping~\cite{maah1,maah2}, even in strictly one-dimensional systems. For an 
AAH model, `energy dependent' mobility edges appear once the second-neighbor hopping is taken into account. It has widespread applications 
and a significant amount of work has already been done~\cite{maah3,maah4}. Witnessing all these facts, researchers have recently been 
focusing on a new type of AAH system, where the site energies vary slowly~\cite{sv1,sv2}, aiming to observe other astonishing phenomena, 
as in such a potential distribution, localization behavior is quite different than the conventional AAH models~\cite{sv3,sv4,sv5,sv6,sv7,sv8}.
Our aim of this work is to explore how a slowly varying disorder potential can influence spin selective electron transfer through a 
magnetic system.   

Spin specific electron transmission is the core part of `spintronics' where both electronic charge and spin are 
involved~\cite{s1,s2,s3,sp1,sp2,sp4}. The use of spintronics is almost everywhere starting from data storage, information processing, sensor
technologies, medical sciences, and to name a few~\cite{sp5,sp6,sp7,sp8,sp9}. The primary concern of spin selectively through any system 
is to separate the up and down spin energy channels, and that can be achieved if the system possesses any spin-dependent scattering
term~\cite{sp10,sp11,fm1,fm2}. We can recall spin-orbit (SO)~\cite{so2,so3,so4,so5} coupling as one the common scattering 
factors, and, in solid-state materials, two different types of SO fields are usually considered that are referred to as Rashba~\cite{rashba} 
and Dresselhaus~\cite{dressel} SO fields. Though these SO fields have several important advantages, they are not particularly suitable 
for favorable responses in many cases, especially when considering a large bias voltage and the possible tuning of spin-specific current 
over a wide range. The reason is that the SO coupling strength is too weak~\cite{smso}, nearly an order of magnitude smaller than the 
electron hopping strength. An alternative prescription is also there which is the use of ferromagnetic (FM) materials~\cite{fm1,fm2}, as 
in such systems we can have strong spin-dependent scattering.
But, ferromagnetic materials are also not perfectly suitable due to some unavoidable reasons those are: poor injection efficiency from 
non-magnetic contacts and high resistivity mismatch at contact points of a nanojunction. Moreover, a possible tuning of spin current 
through a ferromagnetic material by means of external magnetic field is quite challenging as it requires a very high magnetic field 
because of the nano-scale region. All these issues can be circumvented with the use of a magnetic system that has zero net magnetization,
more specifically, an antiferromagnetic (AFM) system~\cite{afm1,afm2}. The inherent stability, high-frequency performance, resilience 
to external magnetic fields, faster switching, and other characteristics make antiferromagnetic materials superior to ferromagnetic
ones~\cite{adafm1,adafm2}. 

Because of these facts, researchers have focused extensively on using AFM systems~\cite{afm3,afm4,afm5} to study spin-dependent 
transport phenomena in recent years. Most models employed the nearest-neighbor TB hopping approximation. However, recent studies on 
helical systems have revealed new
\begin{figure}[ht]
{\centering \resizebox*{3.5cm}{7cm}{\includegraphics{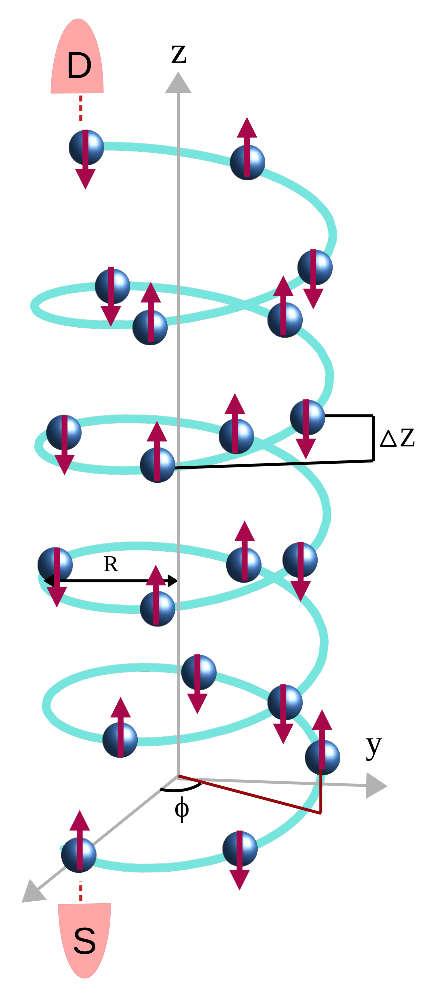}}\par}
\caption{(Color online). Spin filtration setup where an AFM helix with antiparallel arrangement of neighboring magnetic moments is 
coupled to one-dimensional non-magnetic electrodes, source (S) and drain (D). The helix is a right-handed one.}
\label{model}
\end{figure}
pathways, especially due to the unique characteristics of helices, where electron hopping extends beyond nearest-neighbor
sites~\cite{hl3,hl4,hl5,hl6,hlskm1,hlskm2}. This was first highlighted through the experimental observation~\cite{ghl} of spin filtration 
in double-stranded DNA, termed chiral-induced spin selectivity (CISS)~\cite{ciss1,ciss2,ciss3}. Following this discovery, many groups, 
including ours, have explored spin-selective transport in various chiral systems~\cite{sp11,hellight,hl1,hl2}.
For a clean FM helix, one can easily get spin-selective electron transfer. Whereas, for a clean AFM helix, the separation between up and 
down spin energy channels is no longer possible unless we apply a transverse electric field. In the presence of an electric field, the 
symmetry between up and down spin sub-Hamiltonians gets broken, resulting in a finite mismatch between two spin channels, and hence 
spin transport is achieved~\cite{ef1}. In the present communication, we consider a different scenario, where the AFM helix is 
free from an external
electric field but subject to a slowly varying potential. The central focus is to examine the interplay among (i) helicity, (ii) slowly 
varying disorder, and (iii) higher-order electron hopping on spin filtration. This interplay may open a new route in this field and can 
be utilized in different contexts involving electron transport. Describing the quantum system within a TB framework, we compute all the 
results following the well-known Green's function formalism. We find a very high degree of spin filtration, and most importantly, our 
results are valid for a wide range of physical parameters. 

The arrangement of the remaining parts is as follows. In Sec. II we briefly describe the spin filtration setup, TB Hamiltonian, and 
theoretical steps for finding the results. All the results are systematically placed and critically analyzed in Sec. III. Finally, 
a summary is given in Sec. IV. 

\section{Junction setup, TB Hamiltonian and the Method}

This section deals with the nanojunction, its Hamiltonian and the required mathematical steps for calculating the results. Here we describe
them one by one as follows. 

\subsection{Junction setup and TB Hamiltonian}

Let us begin with Fig.~\ref{model} where a right-handed AFM helix is sandwiched between two one-dimensional non-magnetic electrodes. 
The filled colored balls represents the magnetic sites and the arrows indicate the directions of magnetic moments. The neighboring
moments are arranged in the $\pm Z$ directions. The helix geometry is described by three important parameters those are: twisting angle
$\Delta \phi$ which measures the angular twist of one site with respect to the other, stacking distance $\Delta Z$ that gives the spacing
between nearest-neighbor sites, and the radius $R$. These three parameters control the structure and directly influence of the range of 
electron hopping. The most influential parameter is $\Delta Z$. When it is very small i.e., atoms are densely spaced, an electron can 
easily hop from one site to far neighboring sites making the system a `long-range hopping' one, whereas for large $\Delta Z$, electronic
hopping is confined within a few neighboring sites characterizing the helix as a `short-range hopping' system. For theoretical studies 
it is quite convenient to consider both these models by adjusting the physical parameters, and all these models are available in reality.
For instance, DNA and protein molecules are direct examples of SRH and LRH models, and nowadays helical systems can also be designed
in a suitable laboratory.    

We use a TB framework to describe the Hamiltonian of the system. As the nanojunction comprises different parts like AFM helix
and electrodes, we can write the TB Hamiltonian of the full system as a sum
\begin{eqnarray}
H & = & H_{\mbox{\tiny AFH}} + H_{\mbox{\tiny S}} +  H_{\mbox{\tiny D}} + H_{\mbox{\tiny C}}
\label{equ1}
\end{eqnarray}
where, the sub-Hamiltonians are explicitly mentioned as follows.

The sub-Hamiltonian of the helix, $H_{\mbox{\tiny AFH}}$, is written as
\begin{eqnarray}
	H_{\mbox{\tiny AFH}} & = & \sum_n \mbox{\boldmath{$c$}}_n^{\dagger} 
	\left(\mbox{\boldmath{$\epsilon$}}_n-\mbox{\boldmath{$h$}}_n.\mbox{\boldmath{$\sigma$}}\right)\mbox{\boldmath{$c$}}_n
	\nonumber \\
	& + & 
	\sum_{n=1}^{N-1}\sum_{m=1}^{N-n}\left(\mbox{\boldmath{$c$}}_n^{\dagger}\mbox{\boldmath{$t$}}_m \mbox{\boldmath{$c$}}_{n+m} + 
	\mbox{\boldmath{$c$}}_{n+m}^{\dagger}\mbox{\boldmath{$t$}}_m\mbox{\boldmath{$c$}}_{n}\right)
	\label{equ2}
\end{eqnarray} 
where the first term is associated with site energy and the second term is connected to the electron hopping.
All the entries of this sub-Hamiltonian are matrices and they are expressed as
$\mbox{\boldmath{$c$}}_n=\begin{pmatrix}
	c_{n\uparrow} \\
	c_{n\downarrow} 
\end{pmatrix}$,
$\mbox{\boldmath{$c$}}_n^{\dagger}=\begin{pmatrix}
	c_{n\uparrow}^{\dagger} & c_{n\downarrow}^{\dagger} 
\end{pmatrix}$,
$\mbox{\boldmath{$\epsilon$}}_n=\begin{pmatrix}
	\epsilon_n & 0\\
	0 & \epsilon_n
\end{pmatrix}$,
$\mbox{\boldmath{$h$}}_n.\mbox{\boldmath{$\sigma$}}=h_n\begin{pmatrix}
	\cos\theta_n & \sin\theta_n e^{-k\varphi_n}\\
	\sin\theta_n e^{k\varphi_n} & -\cos\theta_n
\end{pmatrix}$
and
$\mbox{\boldmath{$t$}}_m=\begin{pmatrix}
	t_m & 0\\
	0 & t_m
\end{pmatrix}$.
$c_{n\sigma}^{\dagger}$, $c_{n\sigma}$ are the conventional fermionic operators at site $n$ of spin 
$\sigma$ ($\uparrow, \downarrow$). $\epsilon_n$ corresponds to the site energy in the absence of magnetic 
interaction. Under that condition, the site energies for up and down spin electrons are the same, and 
hence we refer them by $\epsilon_n$ (without using any spin index). The term 
$\mbox{\boldmath{$h$}}_n.\mbox{\boldmath{$\sigma$}}$ is responsible for spin-dependent scattering and it
appears due to the interaction of itinerant electrons with local magnetic moments, where $h_n$ measures 
the strength and $\mbox{\boldmath{$\sigma$}}$ is the Pauli spin vector. The factor $h_n$ involves the average 
spin at site $n$ and the spin-moment coupling strength. The orientation of any such magnetic moment is defined
by a polar angle $\theta_n$ and an azimuthal angle $\varphi_n$. The symbol $k$ corresponds to $\sqrt{-1}$.
The parameter $t_m$ denotes the hopping between the sites $n$ and ($n+m$). We scale it with respect
to the nearest-neighbor electron hopping and it is expressed as~\cite{tjlj}
\begin{equation}
t_m=t_1 e^{-(l_m-l_1)/l_c}
\label{equ3}
\end{equation}
where $t_1$ is the nearest-neighbor hopping. In this hopping expression, $l_m$ represents the distance between the sites $n$ and ($n+m$), 
and when $m=1$ we get the nearest-neighbor distance $l_1$. Finding an expression of $l_m$ in terms of the helical parameters is a bit
tricky. From the projection of the helix, first we need to find $R$ and then doing some algebra we arrive at the expression~\cite{tjlj}
\begin{equation}
l_m=\sqrt{[2R\sin(m\Delta\phi/2)]^2+(m\Delta Z)^2}
\label{equ4}.
\end{equation}
From this relation, it is found that as the stacking distance $\Delta Z$ increases, $l_m$ increases rapidly, causing the hopping strength 
to decrease exponentially. So, $\Delta Z$ can play a pivotal role to regulate the electron hopping among different sites. The parameter 
$l_c$ is the decay factor. Instead of exponential decay, one can also think about the hopping in other forms like what are available in 
different other cases, but for helical systems this type of hopping is widely used, and accordingly, we also consider in our present work.  

For our chosen quantum system, we assume site energies ($\epsilon_n$'s) are not uniform, rather they are slowly-varying following a 
cosine form. We can express them as~\cite{sv1}
\begin{equation}
\epsilon_n = w\cos\left(2\pi n^\nu\right)
\label{equ5}
\end{equation}
where $w$ is the disorder strength and $\nu$ is a parameter ranging from $0$ to $1$. For $\nu=0$ or $1$, site energies become identical 
($\epsilon_n=w$ for all $n$) resulting in a perfect system. In such a situation, we can choose $w=0$, without loss of any generality.
Whereas, for any other value of $\nu$ the system becomes disordered with a strength $w$. The change of $\nu$ yields a different 
disordered configuration. Apart from $\nu=0$, when we refer to any other values, it means the values explicitly exclude $\nu=1$, as 
$\nu=0$ and $1$ correspond to the same.  

The sub-Hamiltonians $H_{\mbox{\tiny S}}$ and $H_{\mbox{\tiny D}}$ of non-magnetic source and drain electrodes can be written as
\begin{equation}
	H_{\mbox{\tiny S}} =\sum_{n<1}\mbox{\boldmath{$a$}}_n^{\dagger} 
	\mbox{\boldmath{$\epsilon$}}_0\mbox{\boldmath{$a$}}_n +
	\sum_{n<1} (\mbox{\boldmath{$a$}}_n^{\dagger} \mbox{\boldmath{$t$}}_0 
	\mbox{\boldmath{$a$}}_{n-1}+h.c.)
	\label{equ6}
\end{equation}
and
\begin{equation}
	H_{\mbox{\tiny D}} = \sum_{n>N} \mbox{\boldmath{$b$}}_n^{\dagger} 
	\mbox{\boldmath{$\epsilon$}}_0 \mbox{\boldmath{$b$}}_n +
	\sum_{n>N} (\mbox{\boldmath{$b$}}_n^{\dagger} \mbox{\boldmath{$t$}}_0 
	\mbox{\boldmath{$b$}}_{n+1}+h.c.)
	\label{equ7}.
\end{equation}
The various matrices in these Hamiltonians are\\
$\mbox{\boldmath{$a$}}_n=\begin{pmatrix}
	a_{n\uparrow} \\
	a_{n\downarrow} 
\end{pmatrix}$,
$\mbox{\boldmath{$a$}}_n^{\dagger}=\begin{pmatrix}
	a_{n\uparrow}^{\dagger} & a_{n\downarrow}^{\dagger} 
\end{pmatrix}$, 
$\mbox{\boldmath{$b$}}_n=\begin{pmatrix}
b_{n\uparrow} \\
b_{n\downarrow} 
\end{pmatrix}$,
$\mbox{\boldmath{$b$}}_n^{\dagger}=\begin{pmatrix}
b_{n\uparrow}^{\dagger} & b_{n\downarrow}^{\dagger} 
\end{pmatrix}$,
$\mbox{\boldmath{$\epsilon$}}_{0}=\begin{pmatrix}
	\epsilon_0 & 0\\
	0 & \epsilon_0
\end{pmatrix}$
and
$\mbox{\boldmath{$t$}}_{0}=\begin{pmatrix}
	t_{0} & 0\\
	0 & t_{0}
\end{pmatrix}$\\
where ($a_{n\sigma}^{\dagger}$, $a_{n\sigma}$) and ($b_{n\sigma}^{\dagger}$, $b_{n\sigma}$) represent the fermionic operators employed for 
the source and drain electrodes respectively. $\epsilon_0$ and $t_0$ denote the site energy and nearest-neighbor hopping for the electrodes,
respectively.

The AFM helix is coupled to the source and drain via the coupling strengths $t_S$ and $t_D$ respectively. The coupling Hamiltonian
$H_{\mbox{\tiny C}}$ is written as   
\begin{equation}
	H_{\mbox{\tiny C}} = \mbox{\boldmath{$a$}}_0^\dagger
	\mbox{\boldmath{$t$}}_S \mbox{\boldmath{$c$}}_1 +
	\mbox{\boldmath{$c$}}_N^\dagger \mbox{\boldmath{$t$}}_D
	\mbox{\boldmath{$b$}}_{N+1}+h.c.
	\label{equ8}
\end{equation} 
where
$\mbox{\boldmath{$t$}}_{S}=\begin{pmatrix}
	t_S & 0\\
	0 & t_S
\end{pmatrix}$
and
$\mbox{\boldmath{$t$}}_{D}=\begin{pmatrix}
	t_{D} & 0\\
	0 & t_{D}
\end{pmatrix}$.

\subsection{Theoretical prescription}

Our main focus of this study is to examine the behavior of spin polarization in an AFM helix. We define it in terms of spin-selective 
currents as~\cite{spl1,spl2},
\begin{equation}
P=\frac{I_\uparrow-I_\downarrow}{I_\uparrow+I_\downarrow} \times 100
\label{equ9}
\end{equation}
where $I_\uparrow$ and $I_\downarrow$ are the currents associated with up and down spin electrons respectively. When either of the two
current components is zero, we get a $100\%$ spin polarization and that is always our highest requirement. Exactly opposite to this 
situation, when both the currents are identical, no spin filtration occurs. For the other situation, we get an intermediate value of
$P$, and depending on the dominating current we can have up or down spin polarization. 

The spin current components are calculated following the well-known Landauer-B\"{u}ttiker prescription, where the current is written
in terms of transmission probability~\cite{gf1,gf2}. For a finite bias voltage $V$, the current expression reads
\begin{equation}
	I_{\sigma} = \displaystyle \frac{e}{h} \int T_{\sigma}(E)\,(f_S-f_D)\,dE
	\label{equ10}
\end{equation}
where $e$ and $h$ are the fundamental constants. $f_S$ and $f_D$ are the Fermi-Dirac distribution functions for S and D, respectively,
and they are associated with the electro-chemical potentials $\mu_S=(E_F+eV/2)$ and $\mu_D=(E_F-eV/2)$. $E_F$ denotes the equilibrium
Fermi energy. The system temperature is defined by the parameter $\tau$. In the current expression, $T_{\sigma}(E)$ is the most 
important quantity which is referred to as spin-dependent transmission probability, and we compute it following the standard Green's 
function method. In terms of the retarded and advanced Green's functions, $\mbox{\boldmath{$G$}}^r$ and
$\mbox{\boldmath{$G$}}^a$, different components of spin-dependent transmission probability $T_{\sigma\sigma^\prime}$ are found from 
the expression~\cite{gf1,gf2}
\begin{equation}
	T_{\sigma\sigma^\prime} = \mbox{Tr}\left[\mbox{\boldmath{$\Gamma$}}_S^\sigma \mbox{\boldmath{$G$}}^r 
	\mbox{\boldmath{$\Gamma$}}_D^{\sigma^\prime} \mbox{\boldmath{$G$}}^a\right].
	\label{equ11}
\end{equation} 
From an incident up-spin electron, there are two possibilities at the drain end: up-spin and down-pin components. We refer them as 
$T_{\uparrow\uparrow}$ and $T_{\uparrow\downarrow}$. Similarly, for an incident down-spin electron, we get other two components those
are $T_{\downarrow\uparrow}$ and $T_{\downarrow\downarrow}$. Therefore, the net up and down spin transmission probabilities are:
$T_{\uparrow}=T_{\uparrow\uparrow}+T_{\downarrow\uparrow}$ and $T_{\downarrow}=T_{\downarrow\downarrow}+T_{\uparrow\downarrow}$, 
respectively. In Eq.~\ref{equ11}, $\mbox{\boldmath{$\Gamma$}}_S^{\sigma}$ and $\mbox{\boldmath{$\Gamma$}}_D^{\sigma^{\prime}}$ are 
the coupling matrices for S and D, respectively, and they are expressed in terms of the contact self-energies as
$\mbox{\boldmath{$\Gamma$}}_{S(D)}^{\sigma(\sigma^\prime)}=-2 \mbox{Im} \left[\mbox{\boldmath{$\Sigma$}}_{S(D)}^{\sigma(\sigma^\prime)}
\right]$. The other two factors, $\mbox{\boldmath{$G$}}^r$ and $\mbox{\boldmath{$G$}}^a$, of Eq.~\ref{equ11} are expressed as~\cite{gf1}
\begin{equation}
	\mbox{\boldmath{$G$}}^r=(\mbox{\boldmath{$G$}}^a)^{\dagger}=\left(E \mbox{\boldmath{$I$}} - H_{\mbox{\tiny AFH}} 
	- \boldsymbol{\Sigma}_S^\sigma - \boldsymbol{\Sigma}_D^\sigma\right)^{-1}
	\label{equ12}
\end{equation}
where $\mbox{\boldmath{$I$}}$ is the identity matrix of order $2N\times2N$.

\section{Results and discussion}

As pointed out, our central motivation of this work is to examine the degree of spin filtration in an AFM helix under the influence of 
a slowly varying potential. Based on the above theoretical formulation we have numerically computed different results which essentially
include (i) spin-specific transmission probabilities and (ii) spin filtration under different input conditions. Before delving into the 
results, let us first mention the values of different physical parameters that are remain unchanged throughout the discussion. 

For the side attached electrodes, source and drain, the site energies are fixed at zero and the NNH strength is fixed at $2\,$eV. The
coupling strengths of these electrodes to the helix, $t_S$ and $t_D$, are set at $1\,$eV. In the AFM helix, the NNH strength $t_1$,
slowly-varying cosine modulation strength $w$, and the spin-dependent scattering factor $h$ are chosen at $1\,$eV. As the magnetic moments 
are aligned along the $\pm Z$ directions, the polar angle $\theta_n$ alternates between $0$ and $\pi$ at successive sites, while the 
azimuthal angle $\varphi_n$ is set to $0$ for all magnetic sites for the sake of simplification. The choice of helix parameters is also 
quite relevant. As described earlier, the helix comprises three parameters: twisting angle $\Delta \phi$, stacking distance $\Delta Z$, 
and the radius $R$. Most of the results are computed for the LRH model, unless specified categorically, and for that we choose the parameters
as: $\Delta \phi= 5\pi/9$, $\Delta Z=1.5$\AA, and $R=2.5$\AA. Results with other sets of these parameters are also discussed in the 
appropriate parts. If not stated otherwise, we compute the results at zero temperature considering $N=20$. All the other energies are also 
measured in unit of electron-volt (eV).

Now we present our results one by one in different sub-sections. 

\subsection{Transmission probabilities and spin polarization}

\begin{figure}[ht]
{\centering\resizebox*{8.5cm}{3.7cm}{\includegraphics{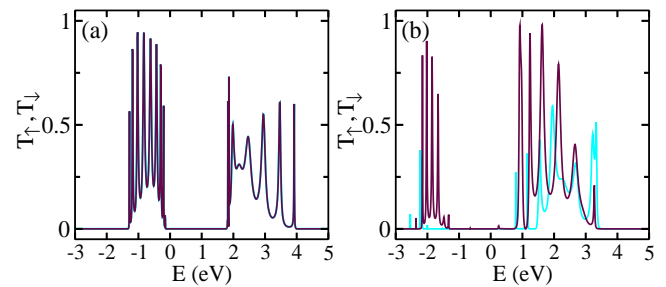}}}
\caption{(Color online). Energy dependent up (cyan) and down (maroon) spin transmission probabilities for an LRH AFM helix, where (a)  and (b) 
correspond to $\nu=0$ and $0.8$, respectively.}
\label{tran}
\end{figure}

We begin with the spin-dependent transmission probabilities, as they serve as the fundamental quantities for determining spin-specific 
currents and, consequently, the degree of spin filtration, represented by the spin polarization coefficient.
As we are dealing with an anti-ferromagnetic system, the first question comes under which condition we can have different spin-dependent
transmission probabilities. Figure~\ref{tran} shows the variations of $T_{\uparrow}$ and $T_{\downarrow}$ as a function of energy $E$ at 
two different values of $\nu$: $\nu=0$ (Fig.~\ref{tran}(a)) and $\nu=0.8$ (Fig.~\ref{tran}(b)). Two colored curves are used for the two
different coefficients. From the spectra it is seen that when $\nu=0$ (ordered case), both up and down spin transmission coefficients 
exactly overlap with each other. 
Whereas, for other value of $\nu$, a finite mismatch between $T_{\uparrow}$ and $T_{\downarrow}$ is 
observed. The underlying mechanism is as follows. For our chosen AFM helix, the neighboring magnetic moments are oriented along $\pm Z$
directions, and hence, the total Hamiltonian $H_{\mbox{\tiny AFH}}$ can be written as a sum of two decoupled sub-Hamiltonians, 
$H_{\mbox{\tiny AFH}\uparrow}$ and $H_{\mbox{\tiny AFH}\downarrow}$, associated with up and down spin electrons, respectively, viz,
$H_{\mbox{\tiny AFH}}$=$H_{\mbox{\tiny AFH}\uparrow}+H_{\mbox{\tiny AFH}\downarrow}$. When $\nu=0$, the system is completely free of 
disorder, and in such a situation $H_{\mbox{\tiny AFH}\uparrow}$ and $H_{\mbox{\tiny AFH}\downarrow}$ are symmetric to each other, resulting
in a same set of energy eigenvalues. Naturally, identical transmission profiles for up and down spin electrons are obtained
(Fig.~\ref{tran}(a)), since the transmission peaks reflect the energy eigenspectra of the conductor clamped between the electrodes.
Once the parameter $\nu$ is set to another value, the helix becomes disordered, and the behavior changes significantly. Here, the
symmetry between $H_{\mbox{\tiny AFH}\uparrow}$ and $H_{\mbox{\tiny AFH}\downarrow}$ is lost due to disorder, and because of that, 
eigenvalues are no longer identical for the up and down spin cases, resulting in a finite mismatch between the spin-specific transmission
spectra (Fig.~\ref{tran}(b)). Usually, in antiferromagnetic systems, this kind of symmetry breaking is achieved by other ways like, applying
transverse electric field, setting non-uniform electron hopping in different segments of a physical systems and the combination of both.
For our chosen case, it is performed by the parameter $\nu$ which plays the central role for spin selective electron transmission, and 
to the best of our concern, such an issue has not been addressed before.     

\begin{figure}[ht]
{\centering\resizebox*{8.2cm}{4cm}{\includegraphics{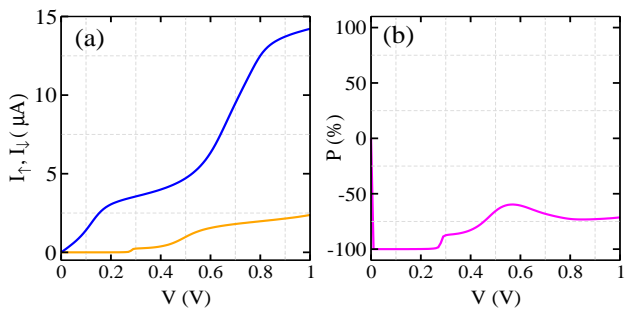}}}
\caption{(Color online). (a) Up (orange line) and down (blue line) spin currents as a function of bias voltage, and (b) corresponding spin 
polarization coefficient, for an LRH AFH helix with $\nu=0.8$. Here we select $E_F=1.3\,$eV.}
\label{current}
\end{figure}

For the ordered AFM helix ($\nu=0$), since $T_{\uparrow}$ and $T_{\downarrow}$ are exactly same, the spin current components associated 
with up and  
down spin electrons will also be identical, and therefore, we cannot expect any spin polarization. It is expected only when there exists a 
finite mismatch between the up and down spin energy channels. As an illustrative example, we consider an AFM helix with a typical $\nu$
($\nu=0.8$) and plot the spin-dependent currents along with the spin polarization coefficient, as shown in Fig.~\ref{current}. The results
are computed selecting $E_F=1.3\,$eV, which is well inside the full energy band. Both the currents increase with the bias voltage, as 
expected, since the current is computed by integrating the transmission profile over a specific energy window associated with the bias.
The natures are quite different for up and down spin cases, reflecting the transmission curves. 
For the down spin case, transmission 
peaks are quite narrow and gapped in the chosen energy window, yielding a step-like behavior in the current (blue line 
in Fig.~\ref{current}(a)). In contrast, for the down spin case, the transmission peaks are broader, leading to a mostly linear current
behavior. The appearance of broadened and narrow transmission peaks is influenced by various factors, with the most significant being 
the conductor-to-electrode coupling and the discrete energy eigenvalues of the conductor. These topics have been discussed in detail 
elsewhere. From the current spectrum it is found that there is a large mismatch between the two current components, reflecting the 
transmission profiles, and accordingly, a high degree of spin polarization is obtained (Fig.~\ref{current}(b)). Up to a moderate voltage
($V \sim 0.2\,$V), almost $100\%$ spin polarization is achieved which is extremely a good signature. Beyond that it decreases, but 
a high degree of spin polarization is still maintained for higher voltage regions. With increasing the bias, more number of up and
down spin transmitting channels are captured, leading to a reduced spin polarization. 
  
\subsection{Robustness of spin polarization: Effects of various parameters}

From the discussion above, it is evident that a high degree of spin polarization can be achieved. This sub-section focuses on the 
robustness of this polarization, specifically examining whether it remains significant across a wide range of different physical 
parameters associated with the system. 

\begin{figure}[ht]
{\centering\resizebox*{8cm}{6cm}{\includegraphics{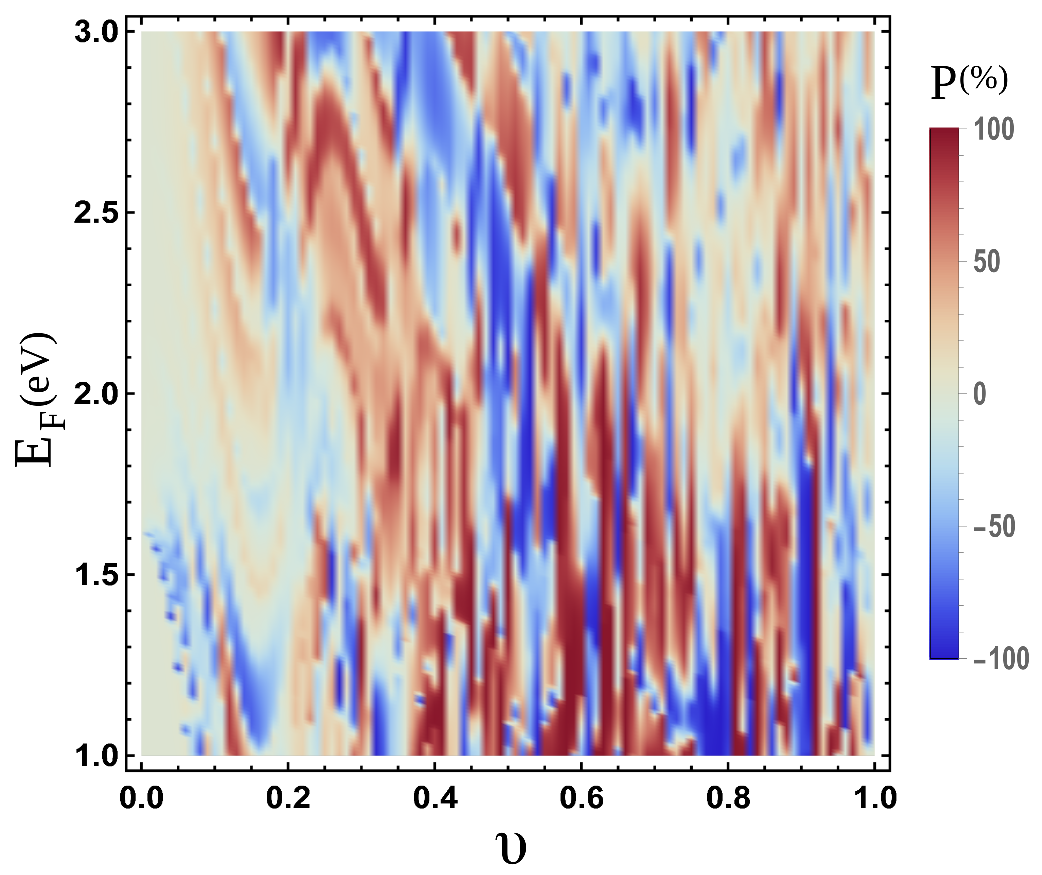}}}
\caption{(Color online). Simultaneous variation of spin polarization with $\nu$ and $E_F$ for an LRH AFM helix when the bias
voltage is set to $0.4\,$V.}
\label{pnuef}
\end{figure}

Figure~\ref{pnuef} shows the simultaneous variation of spin polarization as functions of Fermi energy $E_F$ and the cosine
modulation parameter $\nu$ for an LRH antiferromagnetic helix. Here we set the bias voltage $V$ at $0.4\,$V. Apart from
$\nu=0$ and $1$, the system is disordered for any other values of $\nu$, but for all those $\nu$ values, the response is
not uniform. Within the range $\sim 0.3<\nu<\sim 0.9$, a high degree of spin polarization is obtained. 
\begin{figure}[ht]
{\centering \resizebox*{8cm}{6cm}{\includegraphics{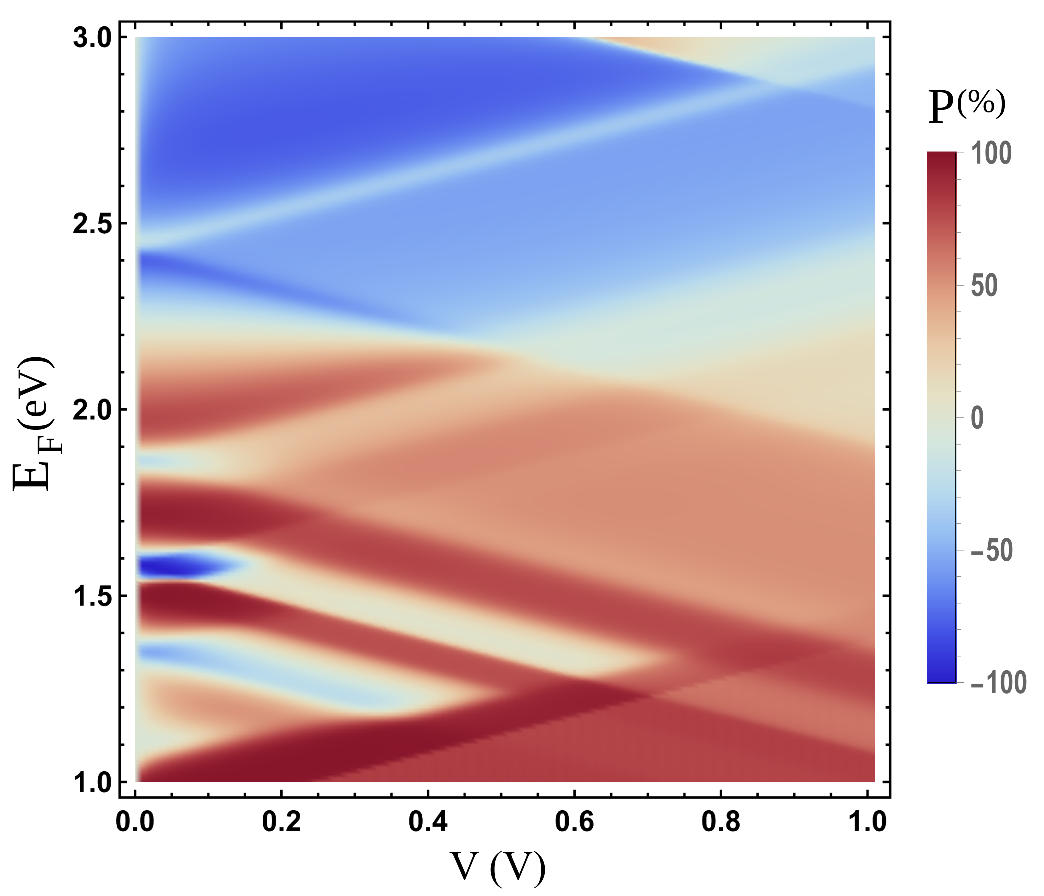}}\par}
\caption{(Color online). Simultaneous variation of spin polarization with $V$ and $E_F$ for an LRH AFM helix with $\nu=0.4$.}
\label{pvef}
\end{figure}
It essentially depends on the mismatch among up and down spin energy channels in presence of disorder. In all these cases, 
the cosine modulation strength $w$ remains the constant, but the configuration is quite important to have a favorable response.
For $\nu=0$ and $1$, no spin polarization is obtained, as discussed earlier.
The choice of Fermi energy, on the other hand, is also important. A large mismatch between up and down spin electrons 
around Fermi energy leads to a high degree of spin polarization. From Fig.~\ref{pnuef} it is noticed that for a wide
range of $E_F$, a favorable response can be achieved, which suggests that a specific choice of $E_F$ is not strictly 
necessary.

In Fig.~\ref{pvef} we show the variation of spin polarization as functions of Fermi energy and the bias voltage. 
From Fig.~\ref{pnuef} it is found that $\nu=0.4$ is a good choice, and hence, in this case we set it for computing
the results. For a broad range of $E_F$ and $V$, the degree of spin polarization is extremely high (close to $100\%$) 
and almost stable. The phase reversal (change of sign) of $P$ occurs upon the change of $E_F$, and one particular
phase persists for a broad range of $E_F$. It clearly suggests that a high degree of up and down spin polarizations
can be substantiated for a wide range of bias window, appropriately choosing the Fermi energy.

\begin{figure}[ht]
{\centering \resizebox*{8cm}{6cm}{\includegraphics{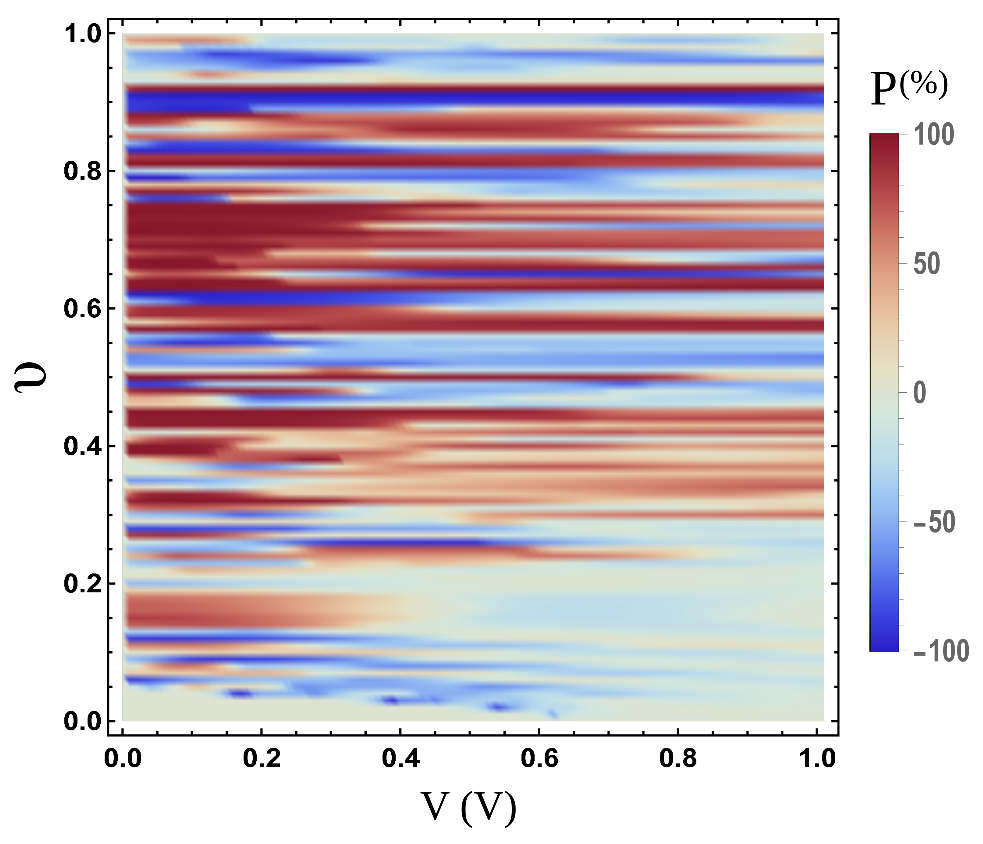}}\par}
\caption{(Color online). Simultaneous variation of spin polarization with $V$ and $\nu$ for an LRH AFM helix when $E_F=1.5\,$eV.}
\label{pvnu}
\end{figure}

In an identical footing, in Fig.~\ref{pvnu} we inspect the simultaneous variation of spin polarization as functions of 
$\nu$ and $V$, keeping the Fermi energy fixed ($E_F=1.5\,$eV). 
Similar to Fig.~\ref{pnuef}, the favorable response is
achieved for $\sim 0.3 <\nu<\sim 0.9$, and that persists almost for the entire chosen bias window.

From all these three density plots viz, Figs.~\ref{pnuef}-\ref{pvnu}, it is clear that a high degree of spin polarization
can be obtained for broad ranges of $\nu$, $V$, and $E_F$, which gives us a confidence that our results can be verified
in a suitable laboratory setup.

\begin{figure}[ht]
{\centering \resizebox*{8cm}{6cm}{\includegraphics{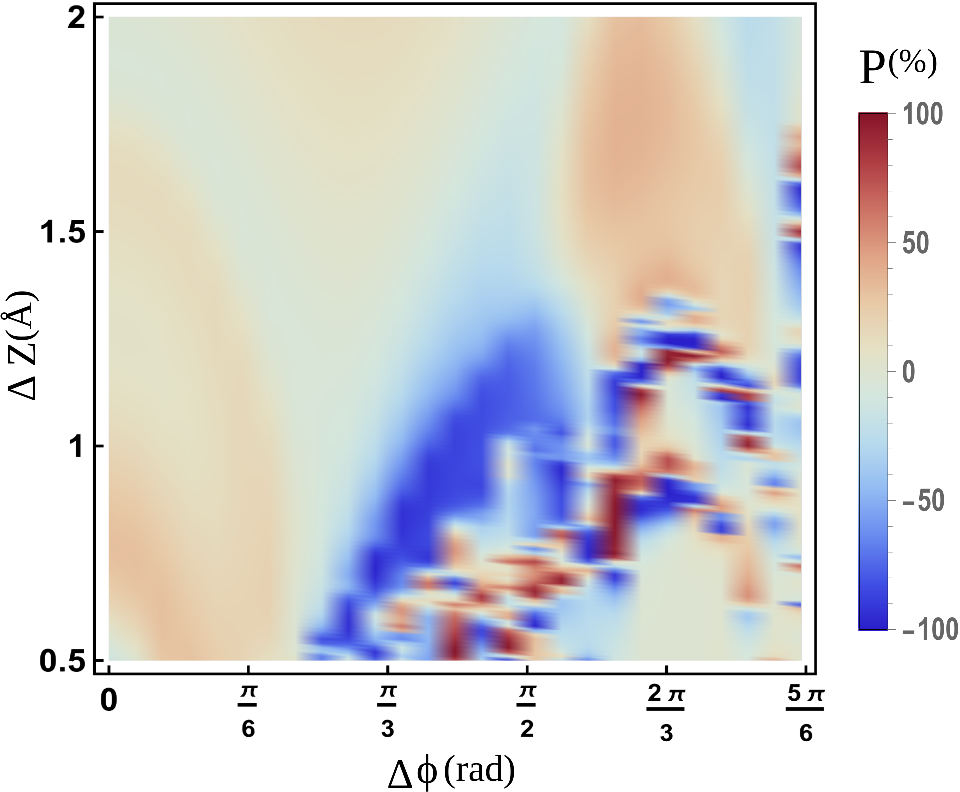}}\par}
\caption{(Color online). Dependence of spin polarization on the structural factors $\Delta Z$ and $\Delta \phi$ of an AFM helix. The other parameters are: $E_F=2\,$eV, $V=0.2\,$V, and $\nu=0.8$.}
\label{pdhdphi}
\end{figure}

Now, we explore the structural dependence, as it is relevant to check how the geometrical conformation affects the spin polarization.
Among the three parameters, $\Delta Z$, $\Delta \phi$, and $R$, we vary the first two keeping the third one as constant and plot the
results in Fig.~\ref{pdhdphi}. Both $\Delta Z$ and $\Delta \phi$ have strong effects on spin polarization. In the absence of any 
twisting ($\Delta \phi=0$), the AFM helical geometry turns into a linear chain, and in this situation, spin polarization becomes 
vanishingly small. Appreciable spin polarizations are obtained for twisted geometries. On the other hand, for a non-zero twisting,
the stacking distance plays its role in other way. The response becomes weaker with increasing the stacking distance. For large 
enough $\Delta Z$, spin polarization again drops almost to zero. These two consequences can be explained as follows.      
As already pointed out earlier, the spin filtration efficiency depends on how much mismatch occurs between up and down spin transmission
\begin{figure}[ht]
{\centering \resizebox*{7cm}{4.5cm}{\includegraphics{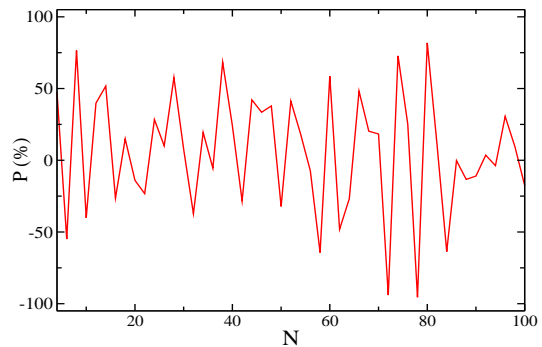}}\par}
\caption{(Color online). Dependence of spin polarization on helix size $N$, with $\Delta N=2$. The other physical parameters are 
$E_F=1.8\,$eV, $\nu=0.8$, and $V=0.2\,$V.}
\label{pns}
\end{figure}
profiles, and that even directly depends on the misalignment of the up and down spin energy channels. The mismatch is originated from 
the symmetry breaking of up and down spin sub-Hamiltonians. For $\nu=0$ and $1$ (perfect cases), the sub-Hamiltonians are symmetric to
each other, regardless of the helicity. For other values of $\nu$, the symmetry is broken, but it becomes effective in presence of 
longer range hopping of electrons. In presence of twisting, the higher order hopping becomes prominent when atoms are closely packed
i.e., $\Delta Z$ is small. Therefore, under those conditions we get favorable responses.

Since we are working in the mesoscopic regime, it is quite essential to check the effect of system size. In Fig.~\ref{pns} we show the
dependence of $P$ on the helix size $N$, by varying $N$ in a large range. We take the interval $\Delta N=2$ to ensure an even number of 
magnetic sites in the helix, thereby maintaining the zero net magnetization condition for our chosen magnetic configuration.
A reasonably large fluctuation (positive $P$ to negative $P$ and vice versa) is noticed, and this is purely due to the effect of 
quantum interference. For large enough system sizes, the fluctuations will be less, but in that regime the degree of spin polarization
might be too low as this is a spin coherence phenomenon. The key insight from the $P$-$N$ profile is that within the coherence regime, 
a high degree of spin polarization can be achieved for different values of $N$. 

\subsection{Role of temperature}

\begin{figure}[ht]
{\centering \resizebox*{7cm}{4.5cm}{\includegraphics{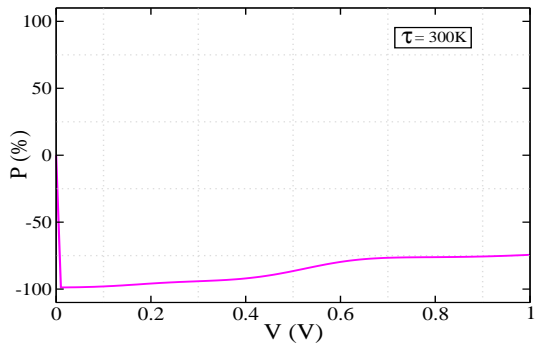}}\par}
\caption{(Color online). Spin polarization as a function of applied bias voltage $V$ for an LRH AFM helix when the system temperature 
is set to $\tau=300\,$K. The other parameters are: $\nu=0.8$ and $E_F=1.25\,$eV.}
\label{pvt}
\end{figure}

In this sub-section we explore the dependence of spin polarization on temperature, which cannot be ignored once we think about 
experimental scenario. In Fig.~\ref{pvt}, we display the variation of spin polarization for an LRH AFM helix setting the 
temperature to $300\,$K. Throughout the presented bias window, spin polarization is highly favorable, and at lower voltages it almost
reaches to $100\%$.

\begin{figure}[ht]
{\centering \resizebox*{7cm}{4.5cm}{\includegraphics{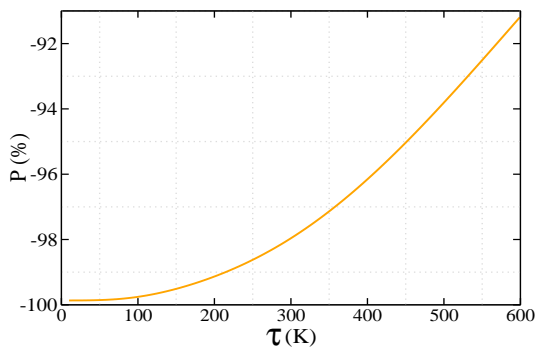}}\par}
\caption{(Color online). Spin polarization $P$ as a function of temperature $\tau$ for an LRH AFM helix. The other parameters 
are: $E_F=1.3\,$eV, $\nu=0.8$, and $V=0.2\,$V.}
\label{pt}
\end{figure}

To have a more clear dependence, in Fig.~\ref{pt} we plot the spin polarization    
as a function of temperature $\tau$ by varying 
it in a wide range, setting the voltage constant. With increasing temperature, the spin polarization decreases, however, even at the
high temperature limit, a reasonably large response is still observed.
The physics behind the reduction of spin polarization with temperature is as follows. At absolute zero temperature, current at a bias 
voltage $V$ is obtained by integrating the transmission profile within the energy window $E_F-eV/2$ to $E_F+eV/2$. So if we set $E_F$
selectively where any of the two spin components dominates, the high degree of spin polarization is expected. For non-zero temperature,
on the other hand, we need to consider the full energy window, and hence, the possibility of getting the contributions from both the 
spin channels increases, which leads to a reduction of spin polarization.   

\section{Closing Remarks}

In this work, we have investigated spin-dependent transport in an antiferromagnetic helix by introducing a slowly varying potential 
instead of an external electric field. We have analyzed how spin polarization depends on various physical parameters and incorporated
temperature effects for a more realistic scenario. Our study has been conducted within a tight-binding model, using the Green’s function
formalism to compute transmission probabilities and the Landauer-Büttiker approach to determine spin-dependent currents. The key findings 
are: \\
$\bullet$ A slowly varying potential breaks system symmetry, leading to a separation of up and down spin transmission probabilities and
resulting in non-zero spin polarization. \\
$\bullet$ By tuning the Fermi energy and $\nu$, almost $100\%$ spin polarization can be achieved.\\
$\bullet$ Spin polarization remains high across a broad range of parameters, showing robustness under different conditions.\\
$\bullet$ The proposed model also exhibits favorable spin-filtering behavior at finite temperatures.

These results suggest that an antiferromagnetic helix with slowly varying site potentials and long-range hopping can serve as an 
efficient spin filter.

\section*{ACKNOWLEDGMENTS}

SS is thankful to DST-SERB, India (File number: PDF/2023/000319) for providing her research fellowship.
DL acknowledges partial financial support from Centers of Excellence with BASAL/ANID financing, AFB220001, CEDENNA.

\section*{DATA AVAILABILITY STATEMENT}

The data that support the findings of this study are available upon reasonable request from the authors.

\section*{DECLARATION}

{\bf Conflict of interest} The authors declare no conflict of interest.

\end{document}